# Approximating Mathematical Semantic Web Services Using Approximation Formulas and Numerical Methods

**Andrei-Horia Mogos\*, Mugurel Ionut Andreica\*\***

*\*Faculty of Automatic Control and Computers,
University Politehnica of Bucharest, Romania (Tel: 0728047112; e-mail:
andrei.mogos@cs.pub.ro)
\*\*Faculty of Automatic Control and Computers,
University Politehnica of Bucharest, Romania (e-mail: mugurel.andreica@cs.pub.ro)*

**Abstract:** Mathematical semantic web services are very useful in practice, but only a small number of research results are reported in this area. In this paper we present a method of obtaining an approximation of a mathematical semantic web service, from its semantic description, using existing mathematical semantic web services, approximation formulas, and numerical methods techniques. We also give a method for automatic comparison of two complexity functions. In addition, we present a method for classifying the numerical methods mathematical semantic web services from a library.

**Keywords:** mathematical semantic web service, web service approximation, complexity functions comparison

## 1. INTRODUCTION

A web service is "a software system designed to support interoperable machine-to-machine interaction over a network" (W3C Working Group, 2004). In order to support important dynamic and automated tasks such as discovery, selection, and composition, web services must be supplied with more semantics. The semantic web, that is a web of machine-processable information, could solve this problem by combining web service technology with the semantic representation of information and, through this, enable automatic and dynamic interaction between software systems (Studer et al., 2007).

Mathematical web services are web services created to solve mathematical problems that can be automatically solved. For example, a mathematical web service can find the solutions of an equation (automatically solvable), or can compute a mathematical expression, or can apply a numerical methods algorithm to a given problem. In order to automate discovery, selection, and composition of mathematical web services, semantic descriptions must be added to these services.

Mathematical semantic web services are very useful in practice, but only a small number of research results are reported in this area. Recently, there are some initiatives to develop languages for the description of mathematical semantic web services. The interest for developing such languages comes from the fact that the mathematical semantic web services have some particularities, and one can develop languages simpler than the existing languages for the semantic web services. As stated in (Mathematical Web Services, 2003) two important initiatives concerned with the semantics of mathematical objects are MathML (MathML, 2003) and OpenMath (OpenMath, 2003), but we need a language for formally describing the semantics of a mathematical operation and these two initiatives are not adequate solutions (Mathematical Web Services, 2003). For our purpose, the most important initiative is the MONET project (MONET Project, 2002). The aim of the MONET project is to demonstrate the applicability of the latest ideas for creating a semantic web to the world of mathematical software, using algorithms to match the characteristics of a problem to the advertised capabilities of available services and then invoking the chosen services through a standard mechanism.

In this paper we present a method of obtaining an approximation of a mathematical semantic web service, from its semantic description, using existing mathematical semantic web services, approximation formulas, and numerical methods techniques. We study only the case when the mathematical semantic web service must compute a mathematical expression. We extend a system described in (Mogos, Florea, 2008). In addition, we discuss an interesting problem: "the automatic comparison of two complexity functions" and give an approximate method for solving the problem. Using this method, we present a way of classifying the numerical methods mathematical semantic web services from a library.

The paper is organized as follows. In section 2, we present the problem we want to solve. Section 3 describes the structure of our system. In section 4, we discuss the problem of comparing two complexity functions. In section 5, we present a method for automatic comparison of two complexity functions. In section 6, we show a method for classifying the numerical methods mathematical semantic web services from a library. Section 7 contains some conclusions and future work.

In the rest of the paper, we will use the abbreviation MSWS for 'mathematical semantic web service'.

## 2. PROBLEM TO DISCUSS

We say that a MSWS $m2$ approximates another MSWS $m1$, if for each input of $m1$, $m2$ gives almost the same output as $m1$. The difference between the outputs of $m1$ and the outputs of $m2$ is controlled by an expression error (Mogos, Florea, 2008).

We understand by semantic description (SD), the semantic information that allows a machine to understand what a service does. We understand by complete description (CD), the entire information used for describing a web service, including its semantic description (Mogos, Florea, 2008).

We want to solve the following problem: given a semantic description of a MSWS, let's call it service $M$, a library of MSWSs, a library of approximation formulas, and a library of MSWSs that use numerical methods algorithms for computations. Our goal is to provide the service M, or at least an approximation of this service.

Since we study only the MSWSs that compute mathematical expressions, the semantic description for such a MSWS can be considered the mathematical expression itself. The mathematical expression gives us sufficient information to construct the MSWS we want as a composed MSWS using simple MSWS that already exist. This is possible because for the mathematical operations used by the mathematical expression we have some priority rules that help us to establish a precise order to compute these operations. A possible approach for this problem is presented in (Mogos, Florea, 2008). In this paper we will extend the system from (Mogos, Florea, 2008).

## 3. SYSTEM STRUCTURE

In this section we present the structure of our system. As one can see in Fig. 1, the system is divided in two parts: the part from Fig. 1a that contains the system structure from (Mogos, Florea, 2008) and the part from Fig. 1b that contains other three modules added to the system by this paper. So, this paper adds three new modules to the system from (Mogos, Florea, 2008) and then the discussion is centred on these three new modules.

The system receives a semantic description of a MSWS and returns a complete description of the MSWS. The Processing Agent uses a recursive algorithm to solve the problem. It communicates with the Decomposition Module and with the Composition Module in order to decompose and to compose the semantic descriptions involved in the problem solving. The Processing Agent also communicates with the Searching Agent. The Searching Agent is responsible with searching MSWSs and approximation formulas in the two libraries; for that reason, it communicates with the MSWS Matching Module and with the Approximation Formulas Matching Module.

The Numerical Methods MSWS Library contains several MSWSs that use numerical methods based algorithms. The Numerical Methods MSWS Matching Module is responsible for finding a MSWS to compute an expression using numerical methods techniques. If this module finds more than one MSWS in the Numerical Methods MSWS Library, then it communicates with the Comparing Agent in order to choose the MSWS (actually, the algorithm) with the lowest temporal complexity.

We consider that every numerical method has a corresponding algorithm. We also consider that, for each algorithm, the complexity is given in terms of $O(f(n))$ where $n$ is the size of the problem, and $f(n)$ is the complexity function related to the algorithm. The Comparing Agent compares two Numerical Methods MSWSs using the complexity functions of the corresponding algorithms. The agent will find the "smallest" MSWS, i.e. the MSWS which has the algorithm with the lowest complexity. The algorithm for comparing two complexity functions is described in section 5.

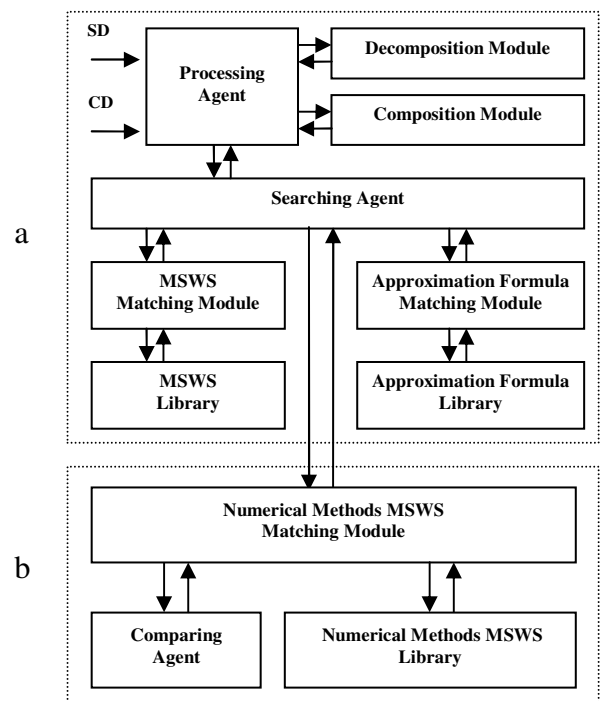

Fig. 1. System structure

The main idea is the following: the system receives a semantic description, i.e. a mathematic expression, and it decomposes the description using an operation priority rule. If all the resulting semantic descriptions correspond to existing MSWSs from the MSWS Library, then the problem is solved. If there is a resulting semantic description without a corresponding MSWS, then the system searches an approximation formula for that description. Due to the fact that the right part of an approximation formula is also a mathematical expression, it can be decomposed in order to compute it using MSWSs from the MSWS Library. If there is no approximation formula for that semantic description, then the system searches a numerical methods MSWS for that semantic description. If there is no numerical methods MSWS that can be used for that semantic description, then the system cannot construct the composed service for the initial semantic description (SD). If there are many numerical methods MSWSs for that semantic description then the MSWS with the lowest temporal complexity algorithm is used. So, in the end, the system constructs a composed MSWS for the initial semantic description, using MSWSs from the MSWS Library and from the Numerical Methods MSWS Library.

Note that the Numerical Methods MSWSs can be used for compute mathematical expressions, so they can be used directly for the MSWS we want to construct, but they can also be used for computing the right part of the approximation formulas. In (Mogos, Florea, 2008), we made the following convention: every MSWS needed for constructing a more complex MSWS necessary for an approximation formula can be found in the MSWS library. For the system described in this paper the following convention is sufficient: every MSWS needed for constructing a more complex MSWS necessary for an approximation formula can be found in the MSWS library or in the Numerical Methods MSWS Library.

## 4. COMPARING TWO COMPLEXITY FUNCTIONS

Comparing the complexity functions of two algorithms can provide useful information regarding the (asymptotic) relative efficiency of the two algorithms and can help us make a more informed decision when multiple algorithms of varying time complexities are available.

Being able to perform such a comparison could also be the building block of a system for automatic classification of algorithms into equivalence classes. In section 6, we will give a method for classifying algorithms.

Let's define the problem in more detail. We have two algorithms for which we know their complexity functions $f_1(n)$ and $f_2(n)$; $n$ is the size of the problem. A complexity function is a function defined on the set of natural numbers with values on the set of positive real numbers. Although the functions are only evaluated at integer values of $n$, we consider that they are defined on the real interval $[1,\infty)$. We denote by $O(f_1(n))$ and $O(f_2(n))$ the complexity classes of the two functions.

We are interested in developing a comparison algorithm *Comp* which takes as inputs the two functions $f_1(n)$ and $f_2(n)$ and returns one of the following results: <1> $O(f_1(n))=O(f_2(n))$, i.e. $f_1(n)$ and $f_2(n)$ are equivalent up to a constant factor; <2> $O(f_1(n)) \subset O(f_2(n))$, i.e. the running time of the second algorithm is larger, as $n$ tends to $+\infty$; <3> $O(f_2(n)) \subset O(f_1(n))$, i.e. the running time of the first algorithm is larger, as $n$ tends to $+\infty$; <4> none of the previous three results holds.

At first, we make the observation that any function (with positive real values) could be the complexity function of an algorithm. Let's consider the following algorithm: *A(n)=if (10≤n≤20) then print the numbers 1,...,n else generate all the permutations with n elements*. $A(n)$ has time complexity $O(n)$ for $10 \leq n \leq 20$ and $O(n!)$ for $1 \leq n \leq 9$ and $n \geq 21$. Thus, the following function, representing the number of basic operations performed by the algorithm, could be the complexity function for A($n$):

$$f(n) = \begin{cases} n, & 10 \leq n \leq 20 \\ n!, & 1 \leq n \leq 9 \quad and \quad n \geq 21 \end{cases} \quad (1)$$

Thus, it is obvious that a complexity function may be discontinuous and even non-increasing (as it would be expected). The general case is the following. Consider any arbitrary function $g(n)$ with positive real values and let's consider that such a function can be computed in $tc(n)$ time. Let $A(n)=\{compute\ x=g(n)\ and\ measure\ t=tc(n);\ for\ i=1\ to\ max\{x-t,0\}\ do:\ perform\ an\ elementary\ operation\}$.

We consider that a complexity function $f(n)$ may have only a finite number of discontinuities. Since the complexity of an algorithm is defined asymptotically, we will only be interested in the last continuous part of $f(n)$ (the part following the last discontinuity). However, this part is not necessarily an increasing function. Consider the function

$$g(n) = 1000 + \left\lfloor 2^{\lceil n \cdot \sin(n \bmod 1000) \rceil} \right\rfloor \quad (2)$$

and use it in the general case presented above. We have $tc(n)=O(n)$ (because the value of $g(n)$ may have $O(n)$ digits). The complexity function $f(n)=max\{g(n),n\}$ is continuous and, for every $n_0 \geq 1$, there exist two values $na$ and $nb$, such that $n_0 \leq na \leq nb$ and $f(na)>f(nb)$. For instance, we can choose $na=n_0 \cdot 1000+1$ and $nb=n_0 \cdot 1000+4=na+3$. We will have:

$$f(na) = \max\{1000 + 2^{na \cdot \sin(1)}, na\},$$
$$f(nb) = \max\{1000, nb\} \quad (3)$$

Since $sin(1) \sim 0.84$ and $sin(4) \sim -0.76$, it is obvious that $1000+2^{0.84 \cdot na}>max\{1000,na+3\}$, for $na \geq 1000$ (which is the case here). Thus, $f(na)>f(nb)$ and $f$ is not an increasing function. Moreover, $f(n)$ is not a decreasing function either. We can choose $na=(n_0+1) \cdot 1000+1$ and $nb=n_0 \cdot 1000+4=na-997$. We have $nb<na$, but $f(nb)<f(na)$ (by the same arguments presented above).

However, we believe that the cases where the last continuous part of a complexity function $f(n)$ is not an increasing function are exceptions. We will now focus on complexity functions with restricted properties in an attempt to develop an efficient function comparison algorithm.

## 5. AN APPROXIMATE METHOD FOR COMPARING TWO COMPLEXITY FUNCTIONS

We will start by considering that the two functions $f_1(n)$ and $f_2(n)$ are *well-behaved* (i.e. they are increasing and continuous and their $p^{th}$ order derivatives are continuous; $0 \leq p \leq pmax$). If we could compute the limit

$$r = \lim_{n \to \infty} \frac{f_1(n)}{f_2(n)} \quad (4)$$

then: if $(r=\infty)$, then the comparison result is <3>; if $(r=0)$, then the result is <2>; if $r$ is a constant value greater than $0$, then the result is <1>; if $r$ does not exist, then the result is <4>. If we know the definitions of $f_1(n)$ and $f_2(n)$, then we could try to use some of the techniques for computing the limits of functions. However, computing limits in an automatic manner has not been successfully achieved so far, even when the actual definition of the functions is known. We will consider that we can evaluate the functions $f$ at any point $n$ (and obtain this value). With this ability, we will perform *black-box testing* on the two functions, disregarding the definition of the two functions (if such a definition is known).

An approximate method to compute the limit is the following: we consider a sequence of values $nv(1)$, …, obtained as follows: $nv(1)$ can be any integer value greater than or equal to $1$; $nv(i \geq 2) = q \cdot nv(i-1)$ ($q \geq 2$). We keep generating values until the absolute difference between the ratios $f_1(nv(i))/f_2(nv(i))$ and $f_1(nv(i+1))/f_2(nv(i+1))$ of every two consecutive values among the last $k \geq 2$ values is at most equal to a fixed small value $\varepsilon > 0$, or until we have generated $L$ values (where $L$ is sufficiently large). If the first condition is met, we will conclude that a bounded limit exists; we just need to decide if this limit is $0$ or strictly positive. Because of this, we repeat the process for the ratio $f_2(n)/f_1(n)$ (with $n=nv(1), nv(2), …$). If we obtain a bounded ratio again, then $r>0$; otherwise, we conclude that $r=0$. If the first condition is not met, we will study the trend of the ratio $f_1(nv(i))/f_2(nv(i))$ for the last (at most) $k$ generated values. If an increasing trend is noticed, then we can conclude that $r=\infty$. This method is not fail-proof. When $L$ values have been generated without reaching a conclusion, the decision made by the method is not too reliable. One of the reasons for this lack of reliability is the following. The two functions, $f_1(n)$ and $f_2(n)$, can intersect in a large number of points (an intersections is a value $nx$, such that $f_1(nx)=f_2(nx)$), and we have no estimate on which is the largest value $nx$ for which this occurs. If, asymptotically, the limit of their ratio is $1$, there may even be an infinite number of intersections.

We will consider here the case when we know that the functions have a finite number of intersections. Thus, the function $dif(x)=f_1(x)-f_2(x)$ has a finite number of roots. Moreover, we will consider that its $p^{th}$ order derivatives ($1 \leq p \leq pmax$) also have a finite number of roots (unless they are the *zero* function). We want to identify the largest such root $nw$, and apply the limit computation method described above, starting from $nv(1)>nw$. There are several methods for finding the largest root. We will start with $xmin=1$ and we want to find the largest root $nw \geq xmin$. We are interested in finding a value $xmax$, such that the interval $[xmin, xmax]$ contains a root. After finding one of the roots $nr$ in this interval, we will set $xmin=nr+1$, and then we will increment $xmin$ by $1$ until $dif(xmin) \neq 0$ and repeat the procedure for the new value of $xmin$, until no more roots can be found. One idea is to start with $xmax=xmin+1$ and repeatedly double the value of $xmax$, until one of the three conditions is met: *1)* $dif(xmax) \cdot dif(xmin)<0$; *2)* $dif'(xmax) \cdot dif'(xmin)<0$ ; *3)* $xmax$ has been doubled for $tmax$ times and none of the conditions *1)* and *2)* has been met. If the condition *1)* is met, we know that $dif(x)$ has a root in the interval $[xmin, xmax]$. We will binary search this root. We will maintain an interval $[a,b]$ during which the root can be found (initially, $a=xmin$ and $b=xmax$), such that $dif(a) \cdot dif(b)<0$. When choosing a value $c$ during the binary search, if $dif(c)=0$, then $c$ is a root; if $dif(c) \cdot dif(b)<0$, we maintain the interval $[c,b]$; otherwise, we maintain the interval $[a,c]$. After finding the root $xr$, we use the procedure described previously for changing the value of $xmin$.

In order to test case *2)*, we need to compute the derivative of the function dif. We will use an approximate method, where $dif'(x)=(dif(x+\varepsilon)-dif(x))/\varepsilon$ (where $\varepsilon>0$ is a very small constant). If we reached case *2)*, then the derivative has a root within the interval $[xmin, xmax]$. We will find the root $xdr$ of the derivative by binary searching it. Afterwards, we repeatedly increase $xmin$, starting from $xdr+1$, until $dif(xmin) \neq 0$.

The method can be generalized as follows. Let's consider that we reached a value of $xmin$. We want to find a value of $xmax$, such that $vdif(p, xmin, xmax)=dif^{(p)}(xmax) \cdot dif^{(p)}(xmin) <0$ ($0 \leq p \leq pmax$); $dif(p)$=the $p^{th}$ derivative of the function *dif*. We repeatedly double $xmax$ and, for each value, we try to find the smallest $p$ for which $vdif(p, xmin, xmax)<0$, binary search the root $xr$ and then set $xmin=xr+1$. Then, while at least one of the values $dif^{(p)}(xmin)=0$, we increment $xmin$ by $1$. As stated previously, we only consider derivatives with a finite number of roots. This excludes the *zero* function, $dif^{(h)}(x)=0$, which is, however, easy to detect. When $xmax$ was doubled too many times, we stop the process and assume that the largest root of the function is smaller than $xmin$. Note that this process takes a finite amount of time, as the function and its derivatives have only a finite number of roots, and we consider only a finite number of derivatives.

Of course, the proposed method is only a heuristic process. However, we are not aware of other approaches for comparing two (general) complexity functions. Nevertheless, the method is applicable on a wide range of common types of functions, e.g. polynomial, poly-logarithmic, exponential, and any of their combinations, as they obey our continuity assumptions.

We have implemented the described method in the Java programming language, due to its in-built facilities to handle *large* numbers and tested it on several standard functions, like polynomials with small degrees (*1*, *2*, or *3*), complexity functions for sorting algorithms (*n·log(n)*) and exponential functions ($2^n$ and *n!*). The classifications were correct every time. Nevertheless, we need to test the method on more complicated complexity functions, in order to better estimate its behaviour and accuracy.

## 6. A METHOD TO ORGANIZE THE LIBRARY OF NUMERICAL METHODS MSWSs

As mentioned in section 3, each Numerical Methods MSWS from the Numerical Methods MSWS Library has a corresponding algorithm. So we can see the Numerical Methods MSWS Library as a library of algorithms. We will use the algorithm *Comp*, described in section 5, for classify the algorithms from a library. But each algorithm has a corresponding complexity function that measures its complexity. So, in fact, we have to classify the complexity functions from a given set of complexity functions. Let be *CF_Set* this set of complexity functions.

We will denote by $R_+$ the set of all positive real numbers and by $N_+$ the set of all natural numbers. We will consider the function $g : N_+ \to R_+$ to be an arbitrary fixed complexity function.

Let's recall the definitions of two important complexity classes (see (Cormen, *et al*., 2001)):

$$O(g(n)) = \{ f : N_+ \to R_+ \mid \exists c \in R_+, \exists n_0 \in N_+ \\ \text{such that } f(n) \leq c \cdot g(n), \forall n \geq n_0 \} \quad (5)$$

$$\Theta(g(n)) = \{ f : N_+ \to R_+ \mid \\ \exists c_1, c_2 \in R_+, \exists n_0 \in N_+ \text{ such that} \\ c_1 \cdot g(n) \leq f(n) \leq c_2 \cdot g(n), \forall n \geq n_0 \} \quad (6)$$

The algorithm *Comp* takes as inputs two complexity functions, $f_1(n)$ and $f_2(n)$, and returns:

- <1>, if $O(f_1(n))=O(f_2(n))$
- <2>, if $O(f_1(n)) \subset O(f_2(n))$
- <3>, if $O(f_2(n)) \subset O(f_1(n))$
- <4>, if the comparison fails

Let be *CF_No* the number of complexity functions from *CF_Set*. We consider *CF_No* sets of complexity functions $C_1, \ldots, C_{CF\_No}$. At the beginning, each of these sets is empty. We associate to each complexity function from *CF_Set* a number from *1* to *CF_No*.

Step 1: Put $f_i(n)$ in $C_i$, for all $i = 1, CF\_No$
Step 2: for all $i = 1, CF\_No - 1$ {

    if $C_i \neq \emptyset$ {
      for all $j = i+1, CF\_No$ {
        r = *Comp* ($f_i(n), f_j(n)$)
        if r = 1 then move $f_j(n)$ in $C_i$ }}}

Step 3: eliminate all the empty sets $C_i$
Step 4: choose for each set a representative function
Step 5: sort increasingly the list of sets, using as order relation the order relation described by *Comp* over the representative functions

The only problem of these steps appears in Step 5, if for two complexity functions, *Comp* returns <4>. We use the following convention: <2> means lower, <3> means greater, and <4> means equal. The fact that <4> means equal, can cause some problems, but in practice this case is not very often.

Consider that the final number of set of complexity functions is *Sets_No*. Suppose that *Comp* is an exact comparison method. Then we have the following property:

$$\forall i \in \{1, \ldots, Sets\_No\}, \forall f(n) \in C_i, \\ C_i = \Theta(f(n)) \cap CF\_Set. \quad (7)$$

This result follows easily from (4) and from the definition of $\Theta(f(n))$. In fact, *Comp* is an approximate comparison method. So, in (7), we have $\approx$ instead of $=$.

Consider that the Numerical Methods MSWS Library is organized using the method presented in this section. In that case, there is no need for the Comparing Agent, because, the Numerical Methods MSWS Matching Module can search in the library starting with the set of Numerical Methods MSWSs (i.e. with the set of corresponding complexity functions) with the index 1, and then increasing the index by 1, until it reaches the MSWS it needs. If there are two MSWSs with similar functionalities in the same set, then they can be seen as identical with respect to our system.

This classification is easy to maintain, even if we add new Numerical Methods MSWSs to our Library. Suppose that we want to add a MSWS with the corresponding complexity function *f(n)*. Then we choose for each set of complexity functions, a representative function, and then compare, using *Comp*, *f(n)* with each representative function. If for one representative function we obtain <1>, then *f(n)* belong in the same set with that function. If for all representative functions we obtain <2>, <3>, or <4>, then we must create a new set for *f(n)*. This new set can be put easily in its place in the list of sets.

If we construct a better complexity functions comparison method, let's call it *Better_Comp* then it is not difficult to obtain a more exact classification, using the already constructed sets. We consider that *Better_Comp* does not contradict *Comp* for the results <1>, <2>, and <3>, but returns less often the result <4> than *Comp*. So, sometimes *Comp* returns <4> and *Better_Comp* returns another result, but every time when *Comp* returns <1> or <2> or <3> *Better_Comp* returns the same result. We have the following property: consider that $f_1(n)$ and $f_2(n)$ are two functions from different sets of complexity functions; if *Better_Comp*( $f_1(n), f_2(n)$ ) = <1> then the corresponding sets should form a single set.

## 7. CONCLUSIONS

In this paper we presented a system that obtain an approximation of a MSWS from the MSWS semantic description, using a library of MSWSs, a library of approximation formulas, and a library of MSWSs for numerical methods. The system extends a system described in (Mogos, Florea, 2008) with a library of MSWSs or numerical methods, a module for searching Numerical Methods MSWSs, and an agent that compares two complexity functions.

An important theoretic part of the paper is the algorithm used by the Comparing Agent to compare two complexity functions. We give an approximate method that works on many usual complexity functions. The functions comparison is necessary to find the algorithm with the lowest temporal complexity for a requested computation.

In section 6, we describe a method for organizing the library of Numerical Methods MSWSs, using the complexity functions comparison method presented section 5. This method is important as an application of the comparison method. It also permit some modifications in the structure of the system; if the library of Numerical Methods MSWSs is organized as explained in section 6, then the Comparing Agent is no longer necessary.

In this paper we considered that all the approximation formulas and all the numerical methods techniques are sufficiently exact such that we don't need to compare their errors in order to get the best approximation of our MSWS. An interesting future work will be to construct a method for comparing the error expressions of two approximation formulas or the error expressions of two numerical methods techniques.

Another possible future work will be to analyze error propagation, because depending of the form of the entire expression, the use of several approximation formulas, and several numerical methods techniques for compute it, may lead in the end to a result significantly different from the real one.